\newif\ifEPLclass
\EPLclassfalse
\ifEPLclass
  \documentclass[doublecol]{epl2}
\else
  \documentclass[twocolumn,10pt,a4paper]{article}
\fi

\usepackage{amsmath}
\usepackage{amssymb}
\usepackage{latexsym}
\usepackage{bm}

\usepackage{xcolor}
\usepackage{graphicx}

\ifEPLclass\else
  \usepackage[margin=2cm]{geometry}
  \usepackage{flushend}
  
  \makeatletter
  \newcommand{\inst}[1]{\textsuperscript{#1}}
  \def\and{, }
  \def\EPL@institute{}
  \def\EPL@abstract{}
  \def\EPL@pacs{}
  \newcommand{\institute}[1]{\gdef\EPL@institute{#1}}
  \newcommand{\shorttitle}[1]{}
  \newcommand{\shortauthor}[1]{}
  \newcommand{\pacs}[2]{\g@addto@macro\EPL@pacs{\par\noindent PACS #1 -- #2}}
  \renewcommand{\abstract}[1]{\gdef\EPL@abstract{#1}}
  
  \renewcommand\section{\@startsection{section}{1}{\z@}%
    {-3.5ex \@plus -1ex \@minus -.2ex}{2.3ex \@plus.2ex}%
    {\normalfont\Large\bfseries\raggedright}}
  \renewcommand\subsection{\@startsection{subsection}{2}{\z@}%
    {-3.25ex\@plus -1ex \@minus -.2ex}{1.5ex \@plus .2ex}%
    {\normalfont\large\bfseries\raggedright}}
  \renewcommand{\maketitle}{%
    \twocolumn[{\centering
      {\LARGE\bfseries \@title\par}\vspace{1em}
      {\large \@author\par}\vspace{0.6em}
      {\small \EPL@institute\par}\vspace{1em}
      \begin{minipage}{0.88\textwidth}\small
        \textbf{Abstract} -- \EPL@abstract\par\vspace{0.5em}
        {\footnotesize\EPL@pacs\par}
      \end{minipage}\par\vspace{1.5em}}]}
  \makeatother
\fi

\title{Constraining the generalized uncertainty principle with neutron interferometry}
\shorttitle{Constraining the GUP with neutron interferometry}

\author{Fabiano Feleppa\inst{1} \and Hooman Moradpour\inst{2} \and Christian Corda\inst{3} \and Sarah Aghababaei\inst{4}}
\shortauthor{F. Feleppa et al.}

\institute{
  \inst{1} Institute for Theoretical Physics, Utrecht University, Princetonplein 5, 3584 CC Utrecht, The Netherlands\\
  \inst{2} Research Institute for Astronomy and Astrophysics of Maragha (RIAAM), University of Maragheh, P.O. Box 55136-553, Maragheh, Iran\\
  \inst{3} International Institute for Applicable Mathematics and Information Sciences (IIAMIS), B.M. Birla Science Centre, India\\
  \inst{4} Department of Physics, Faculty of Sciences, University of Sistan and Baluchestan, Zahedan, Iran
}
\pacs{04.60.Bc}{Phenomenology of quantum gravity}
\pacs{03.65.Vf}{Phases: geometric}
\pacs{03.75.-b}{Matter waves}

\abstract{A non-zero minimal length arises in various approaches to quantum gravity and leads to a generalized uncertainty principle (GUP). We analyze its effects on neutron interferometry within a phenomenological model. The GUP correction to the Sagnac phase of a small horizontal interferometer depends on the neutron mass and velocity, while the gravitationally induced (COW) phase acquires a correction at fixed de Broglie wavelength. Special- and general-relativistic corrections are also discussed. We use these results to obtain sensitivity estimates for the dimensionless GUP parameter.}

\begin{document}

\maketitle

\section{Introduction}
The Heisenberg uncertainty principle (HUP) allows the uncertainty in position to be made arbitrarily small, at the price of an arbitrarily large uncertainty in momentum. This conclusion becomes questionable once gravity is taken into account~\cite{grav}. Moreover, various approaches to quantum gravity suggest generalized versions of the HUP, known as the GUP:
\begin{equation}\label{guprelation}
\Delta X\Delta P\geq\frac{\hbar}{2}[1+\beta(\Delta P)^2+\cdots],
\end{equation}
where $\beta$ denotes the GUP parameter~\cite{prdgup, GUPworks}. The above relation implies a non-zero lower bound on $\Delta X$, which is of the order of the Planck length if $\beta$ is of the order of $(M_{p}c)^{-2}$, where $M_{p}$ is the Planck mass; this is also supported by thought experiments~\cite{scardigli1}.
Such a generalization has implications for a wide range of physical systems~\cite{GUPwork2}. From the theoretical point of view, the dimensionless GUP parameter $\beta_{0}\equiv\beta M_{p}^{2}c^{2}$ is usually assumed to be of order unity~\cite{GUPth}; however, this choice makes quantum-gravity effects too small to be measured. When this parameter is allowed to vary, current experiments set upper bounds on $\beta_{0}$ that are much larger than unity; values of $\beta_{0}\gg 1$ compatible with these bounds could signal the existence of a new intermediate length scale~\cite{boundsGUP}.

Several decades ago, Werner and co-workers showed that gravitational~\cite{sag1, sag2} and rotational~\cite{NSagnac} effects can be observed in neutron interference experiments performed on the Earth's surface~\cite{sag3, sag4}. The COW experiment was the first observation of a quantum-mechanical interference effect induced by gravity: the measured phase shift depends on both the gravitational acceleration $g$ and Planck's constant $\hbar$. The two experiments are referred to hereafter as the COW experiment and the neutron Sagnac effect, respectively. The Sagnac effect has been studied extensively because of its importance for fundamental physics (see, for instance,~\cite{sagnac1, sagnac2}), and the effects of modified uncertainty relations on interferometric phases, including the Sagnac phase, have been considered in Ref.~\cite{Mirza}.

In the present work, we show how a non-zero minimal length affects the phase shifts measured in these experiments; in particular, the corrections depend on the mass and velocity of the neutron. Starting from these corrections, we then estimate the sensitivity to the dimensionless parameter $\beta_{0}$.

The paper is organized as follows. In the next section, we briefly recall some important facts about the GUP and the deformed algebra used in this work. Then, we study the GUP-deformed COW effect. Next, we investigate the modification induced by the GUP on the neutron Sagnac phase shift, in both the non-relativistic and the special-relativistic regimes; a general-relativistic application involving the Kerr metric is also considered. Finally, in the last section, we constrain the GUP parameter and summarize our results.

\section{Deformed algebra}
Investigations in string theory and quantum gravity (see, e.g.,~\cite{Gup}) lead to the GUP:
\begin{eqnarray}
\Delta X \geq \frac{\hbar}{2}\left(\frac{1}{\Delta P} + \beta \Delta P\right),
\label{GUP}
\end{eqnarray}
from which follows the existence of a minimal length $\Delta X_{min}=\hbar \sqrt{\beta}$, which is of the order of the Planck length $l_{p}=\sqrt{\hbar G/c^{3}}$ if $\beta_{0}=\beta M_{p}^{2}c^{2}$ is of order unity. It has been shown that a minimal length can be obtained from a small quadratic deformation of the Heisenberg algebra~\cite{prdgup}:
\begin{eqnarray}\label{Diraccom}
[X,P] = i\hbar (1+\beta P^{2}).
\end{eqnarray}
In the classical limit ($\hbar \rightarrow 0$), the quantum-mechanical commutator is replaced by the Poisson bracket of the corresponding classical variables,
\begin{eqnarray}
\frac{1}{i\hbar}[X,P] \rightarrow \left \{ X,P \right \},
\end{eqnarray}
which, in the deformed case~(\ref{Diraccom}), gives
\begin{eqnarray}
\left \{ X,P \right \} = 1+\beta P^{2}.
\label{poison}
\end{eqnarray}
Eq.~(\ref{poison}) shows that the Poisson brackets are deformed in the same way as the quantum commutator. As a consequence, the equation of motion of a test particle in a gravitational field depends on the mass (and velocity) of the particle, which signals a violation of the equivalence principle (EP)~\cite{EP}. An interesting discussion of GUP-deformed Poisson brackets can be found in Ref.~\cite{Cas}. Whether the GUP modifications of the Sagnac effect can be disentangled from a (possible) violation of the EP is beyond the scope of this paper.

The fact that the GUP can be derived from a deformed Heisenberg algebra makes it possible to study the influence of a minimal length on physical systems, at both the quantum and the classical level. In order to study the GUP effects on the systems considered here, we introduce the following deformed algebra:
\begin{align}
[X_{i},P_{j}]&=i\hbar\sqrt{1+\beta P^{2}}\left(\delta_{ij}+\beta P_{i}P_{j}\right),\label{algebra}\\
[X_{i}, X_{j}]&=[P_{i}, P_{j}]=0,
\end{align}
which can be obtained using the representation
\begin{equation}\label{correp}
\mathbf{X}=\mathbf{x},\qquad
\mathbf{P}=\frac{\mathbf{p}}{\sqrt{1-\beta \mathbf{p}^2}},
\end{equation}
where $\mathbf{x} = (x_1,x_2,x_3)$ and $\mathbf{p} = (p_1,p_2,p_3)$ are the canonical position and momentum, which satisfy the canonical commutation relations
\begin{equation*}
[x_{i},p_{j}]=i\hbar \delta_{ij}, \quad [x_{i},x_{j}]=[p_{i},p_{j}]=0.
\end{equation*}
Representations of this type, in which the position operator is undeformed and the physical momentum is a function of the canonical one, are also considered in Refs.~\cite{GUPworks, rep}. Note that, in one dimension, Eq.~(\ref{algebra}) becomes $[X,P]=i\hbar(1+\beta P^{2})^{3/2}=i\hbar[1+\tfrac{3}{2}\beta P^{2}+\mathcal{O}(\beta^{2})]$. Therefore, at first order, $\beta$ is $2/3$ of the parameter in Eq.~(\ref{Diraccom}); this factor preserves the estimated orders of magnitude of the bounds derived below. For $\beta>0$, the corresponding uncertainty relation gives $\Delta X\geq\hbar[1+\beta(\Delta P)^{2}]^{3/2}/(2\Delta P)$. Minimizing the right-hand side over $\Delta P$ yields the state-independent lower bound $\Delta X\geq(3\sqrt{3}/4)\,\hbar\sqrt{\beta}$.

\section{GUP-deformed COW effect}
We first neglect the Earth's rotation and use the local linear approximation to its gravitational potential.
The experiments of Colella, Overhauser, and Werner, commonly referred to as the COW experiments, provided the first observation of gravitationally induced quantum interference. The gravitational field is treated in Newtonian terms, but the formula for the phase shift contains both the gravitational acceleration $g$ and Planck's constant $\hbar$. The experimental set-up is described in detail in Ref.~\cite{sag2}; it is based on splitting the neutron beam by Bragg diffraction from perfect crystals, as first implemented for X-rays by Bonse and Hart~\cite{Bonse}. The two paths form a closed loop, which we approximate by a rectangle of centimetre dimensions. The neutrons enter at the bottom-left corner, where the beam splits into two, so that the two beams travel at different heights in the Earth's gravitational field, with different velocities. The beams then recombine at the top-right corner, where interference takes place. Let us briefly recall a simple derivation of the COW effect. The spatial part of a plane matter wave (a neutron beam in this case) is $e^{i\mathbf{k}\cdot \mathbf{r}}$, where $\mathbf{k}$ is the wave vector and $k \equiv |\mathbf{k}|=2\pi/\lambda$ is the wave number, with $\lambda=2\pi\hbar/p$ the de Broglie wavelength ($p$ is the canonical momentum). The phase accumulated by the neutrons along a given path is then written as follows:
\begin{equation}
\phi(\mathbf{r})=\frac{1}{\hbar} \int_{\mathbf{r}_{0}}^{\mathbf{r}} \mathbf{p} \cdot d \mathbf{r}.
\end{equation}
The phase difference is then
\begin{equation}
\Delta \phi=\frac{1}{\hbar} \int_{\mathbf{r}_{0}}^{\mathbf{r}}\left(\mathbf{p}_{\mathrm{1}}-\mathbf{p}_{\mathrm{2}}\right) \cdot d \mathbf{r},
\end{equation}
where the subscripts $1$ and $2$ refer to the lower and upper paths, respectively. The only non-vanishing contributions to the phase difference come from the horizontal segments of the paths; writing $p_{1}=mv$ and $p_{2}=mv'$, where $v$ and $v'$ are the speeds of the neutron along the two horizontal segments, we get
\begin{equation}
\Delta \phi=\frac{1}{\hbar} m(v-v') L,
\end{equation}
where $L$ is the length of the interferometer. Energy conservation gives
\begin{equation}
\frac{1}{2} m v'^{2}=\frac{1}{2} m v^{2}-m g H,
\end{equation}
where $g$ is the gravitational acceleration and $H$ the
height of the interferometer. Since $gH\ll v^{2}$ (thermal neutrons were used in the COW experiment), we have $v-v' \simeq gH/v$, and we arrive at the final result
\begin{equation}\label{COWstd}
\Delta \phi_{COW}=\frac{m g A}{\hbar v}=\frac{m^{2}gA\lambda}{2\pi\hbar^{2}},
\end{equation}
where $A=LH$ is the area of the interferometer.

The above results can also be obtained from the Lagrangian
\begin{equation}\label{lagrangian}
\mathcal{L} = \frac{1}{2}m\dot{\mathbf{r}}^{2}+m\mathbf{g}\cdot \mathbf{r},
\end{equation}
where the vector $\mathbf{g}$ points downwards, so that the potential energy is $U(\mathbf{r})=-m\mathbf{g}\cdot\mathbf{r}$, and $\mathbf{p}=\partial \mathcal{L}/\partial\dot{\mathbf{r}}=m\dot{\mathbf{r}}$.

In order to analyze the influence of a minimal length on the COW effect, it is commonly assumed that the deformed Hamiltonian has the same functional form as the undeformed one, with the canonical momentum replaced by the deformed one. Using the $P(p)$ relation in~(\ref{correp}), we then obtain
\begin{align}\label{H1}
H^{GUP}_{COW}&=\frac{P^{2}}{2m}+U(\mathbf{r})\nonumber\\
&=\frac{p^{2}}{2m(1-\beta p^{2})}+ U(\mathbf{r})\nonumber\\
&=\frac{p^{2}}{2m}+ \beta\frac{p^{4}}{2m}+U(\mathbf{r})+ \mathcal{O}\left(\beta^{2}\right)\nonumber\\
&\simeq H_{COW}+\beta\frac{p^{4}}{2m}.
\end{align}
For a monochromatic beam undergoing elastic diffraction in a stationary interferometer, both paths share the same conserved total energy $E$. The kinetic energy in Eq.~(\ref{H1}) is therefore lower by $mgH$ on the upper horizontal segment than on the lower one. To first order in $mgH$, the canonical momenta $p_{1}$ and $p_{2}$ on the lower and upper segments, respectively, satisfy
\begin{equation}
p_{1}-p_{2}\simeq\frac{mgH}{v},
\end{equation}
where $v=\partial H^{GUP}_{COW}/\partial p=p/[m(1-\beta p^{2})^{2}]$ is the velocity of the neutron, given by Hamilton's equation. To first order, $v=(p/m)(1+2\beta p^{2})+\mathcal{O}(\beta^{2})$. Therefore, we obtain
\begin{eqnarray}
\Delta \phi_{COW}^{GUP}=\frac{1}{\hbar}\oint \mathbf{p} \cdot d\mathbf{r}=\frac{mgA}{\hbar v}.
\label{COWv}
\end{eqnarray}
For the non-rotating COW calculation, Hamilton's equation $v=\partial H/\partial p$ yields the standard velocity form to all orders in $\beta$ and first order in $mgH$. In COW-type experiments, Bragg diffraction selects the de Broglie wavelength $\lambda=2\pi\hbar/p$, i.e., the canonical momentum $p$. Energy is conserved along both arms, whereas the comparison between theories is made at fixed incident wavelength. Expressed in terms of $\lambda$, the phase shift reads
\begin{eqnarray}
\Delta \phi_{COW}^{GUP}=\frac{m^{2}gA\lambda}{2\pi\hbar^{2}}\left[1-2\beta\left(\frac{2\pi\hbar}{\lambda}\right)^{2}\right]+\mathcal{O}(\beta^{2}).
\label{COWGUP}
\end{eqnarray}
In the limit $\beta\rightarrow 0$, we recover the standard result. Therefore, at fixed wavelength, the relative deviation induced by the GUP on the COW effect is
\begin{eqnarray}
\left\lvert\frac{\Delta \phi_{COW}^{GUP}-\Delta \phi_{COW}}{\Delta \phi_{COW}}\right\rvert=2\beta p^{2}\simeq 2\beta m^{2}v^{2},
\label{dCOW}
\end{eqnarray}
where $\Delta \phi_{COW}=m^{2}gA\lambda/(2\pi\hbar^{2})$ denotes the standard phase shift; above, we have used $p=mv+\mathcal{O}(\beta)$.

\section{GUP-deformed neutron Sagnac effect}
In 1975, Page observed that the rotation of the Earth induces corrections to the phase shift caused by the Earth's Newtonian potential~\cite{Page}; for typical interferometers, these corrections are about $10^{-2}$ of the COW phase~\cite{Ryder2017}. This phenomenon is commonly regarded as the counterpart of the Sagnac effect for matter waves. The experiment was performed in 1979 by Werner, Staudenmann, and Colella~\cite{NSagnac}. In the rotating frame of the Earth, the Hamiltonian governing the motion of the neutron contains a third term in addition to the kinetic and gravitational potential energies~\cite{Ryder2017}, and the momentum becomes
\begin{equation}
\mathbf{p}=m \mathbf{\dot{r}}+m \bm{\omega} \times \mathbf{r}.
\end{equation}
The phase arising from the term in $\omega$ is
\begin{equation}
\Delta \alpha_{S}=\frac{1}{\hbar} \oint m[\bm{\omega} \times \mathbf{r}] \cdot d \mathbf{r}=\frac{2 m \bm{\omega} \cdot \mathbf{A}}{\hbar}.
\end{equation}
Explicitly, the Hamiltonian reads
\begin{equation}\label{eq:Sagnac}
H_{S} = \frac{p^{2}}{2m} +m\phi(r) - \bm{\omega} \cdot \mathbf{L},
\end{equation}
where $S$ denotes the Sagnac effect, $\mathbf{L}=\mathbf{r}\times\mathbf{p}$, and $\phi(r)=-GM/r$. Here $\mathbf{r}$ is measured from the centre of the Earth in co-rotating coordinates and $\mathbf{p}$ is the corresponding canonical momentum. We retain the phenomenological prescription of deforming this momentum in both the kinetic and rotational terms. For the Sagnac results below, we consider a small horizontal interferometer, with $\mathbf{A}$ parallel to the local radial direction, so that the potential and the momentum magnitude at $\omega=0$ are constant along the paths to leading order in the apparatus size. Eq.~(\ref{eq:Sagnac}) can also be written as
\begin{equation}
H_{S} = \frac{p^{2}}{2m} +m\phi(r) - \mathbf{p} \cdot (\bm{\omega} \times \mathbf{r}).
\end{equation}
By replacing the canonical variables with the deformed ones, we have
\begin{align}
H^{GUP}_{S} &= \frac{P^{2}}{2m} +m\phi(r) - \mathbf{P} \cdot (\bm{\omega} \times \mathbf{r})
\nonumber\\
&= \frac{1}{2m}\frac{p^{2}}{1-\beta p^{2}} +m\phi(r) - \frac{\mathbf{p}\cdot (\bm{\omega} \times \mathbf{r})}{\sqrt{1-\beta p^{2}}}.\label{eq:HSGUP}
\end{align}
To first order in the deformation parameter $\beta$, the Hamiltonian reads
\begin{align}
H^{GUP}_{S} &= \frac{p^{2}}{2m} + \frac{\beta}{2m}p^{4} +m\phi(r)
\nonumber\\
& \hspace{0.4cm} - \mathbf{p}\cdot (\bm{\omega} \times \mathbf{r})\left(1+\frac{1}{2}\beta p^{2}\right) + \mathcal{O}\left(\beta^{2}\right).
\end{align}
Hamilton's equation gives, to first order in $\beta$,
\begin{align}
\dot{\mathbf{r}}&=\frac{\mathbf{p}}{m}(1+2\beta p^{2})
-\left(1+\frac12\beta p^{2}\right)(\bm{\omega}\times\mathbf{r})\nonumber\\
&\quad-\beta\mathbf{p}\left[\mathbf{p}\cdot(\bm{\omega}\times\mathbf{r})\right].
\end{align}
For a monochromatic beam undergoing elastic diffraction from crystals at rest in the rotating frame, both arms share the same conserved energy $E$. Hereafter, a subscript $0$ on momentum or velocity denotes its value at $\omega=0$, at the same energy $E$. We write $\mathbf{p}=\mathbf{p}_{0}+\delta\mathbf{p}$, with $p_0=|\mathbf{p}_0|$. At fixed energy, rotation changes the canonical momentum component along the path direction $\hat{\mathbf{n}}$ by
\begin{equation}
\delta p\equiv\hat{\mathbf{n}}\cdot\delta\mathbf{p}
=m\,\hat{\mathbf{n}}\cdot(\bm{\omega}\times\mathbf{r})
\left(1-\frac32\beta p_0^{2}\right).
\end{equation}
Here and below, rotational phases are evaluated to first order in $\omega$. Evaluating their coefficients at $\omega=0$ neglects only terms of order $\omega^2$. Subtracting the phase at $\omega=0$ and using the horizontal geometry specified above gives
\begin{equation}\label{eq:GUPsag}
\Delta\alpha_{S}^{GUP}=\frac{1}{\hbar}\oint\delta\mathbf{p}\cdot d\mathbf{r}
=\Delta\alpha_{S}\left(1-\frac32\beta p_0^{2}\right).
\end{equation}
The undeformed phase is recovered for $\beta\to0$. Since $p_0=mv_0+\mathcal{O}(\beta)$, the leading relative deviation is
\begin{equation}\label{dNR}
\left|\frac{\Delta\alpha_{S}^{GUP}-\Delta\alpha_{S}}{\Delta\alpha_{S}}\right|
=\frac32\beta m^{2}v_0^{2}.
\end{equation}
Here $v_0$ is the neutron speed at $\omega=0$, and the reference phase must dominate the neglected contributions. The expansion assumes $\beta p_0^{2}\ll1$, $\ell\ll R$, $g\ell\ll v_0^{2}$ and $|\bm{\omega}\times\mathbf{r}|\ll v_0$ along the paths, where $\ell$ is the interferometer size and $R$ the Earth's radius. The last condition concerns the laboratory's speed around the Earth's rotation axis.
Let us now include the special-relativistic kinetic energy, retaining the same Earth-centred prescription and horizontal geometry. For $\beta=0$, the zero-rotation kinetic energy and momentum are $(\gamma_0-1)mc^{2}$ and $\mathbf{p}_0=\gamma_0m\mathbf{v}_0$, respectively~\cite{NRform}, where $\gamma_0=(1-v_0^{2}/c^{2})^{-1/2}$. As a function of the canonical momentum magnitude $p$, the kinetic energy expands as
\begin{align}
\sqrt{p^{2}c^{2} + m^{2}c^{4}} - mc^{2} &= \frac{p^{2}}{2m} - \frac{p^{4}}{8m^{3}c^{2}}\nonumber\\
&\hspace{0.4cm}+ \mathcal{O}\left(1/c^{4}\right).
\end{align}
The corresponding Hamiltonian in the rotating frame is written as
\begin{align}
H_{R,S}&= \sqrt{p^{2}c^{2}+m^{2}c^{4}}-mc^{2}\nonumber\\
&\quad +m\phi(r)-\bm{\omega}\cdot\mathbf{L}\nonumber\\
&=\frac{p^{2}}{2m} - \frac{p^{4}}{8m^{3}c^{2}} +m\phi(r)\nonumber\\
&\hspace{0.4cm}- \mathbf{p} \cdot (\bm{\omega \times \mathbf{r}}) +
\mathcal{O}\left(1/c^{4}\right),
\end{align}
where $R$ and $S$ stand for relativistic and Sagnac, respectively (we consider relativistic effects up to order $1/c^{2}$). The rotating-frame transformation gives the term $-\bm{\omega}\cdot\mathbf{L}$. The Lorentz factor in the Sagnac phase arises from the kinetic term:~at fixed energy, $\delta p=\gamma_0m\,\hat{\mathbf{n}}\cdot(\bm{\omega}\times\mathbf{r})$, giving $\Delta\alpha_{R,S}=\gamma_0\Delta\alpha_{S}$ for constant $p_0$ along the loop, in agreement with Ref.~\cite{Ryder2017}.

Again, by replacing the canonical variables with the deformed ones and
by using the $P(p)$ relation, we can write the Hamiltonian as a function
of the canonical variables $p$ and $r$:
\begin{align}
H^{GUP}_{R, S} &= \sqrt{P^{2}c^{2}+m^{2}c^{4}}-mc^{2}\nonumber\\
&\quad +m\phi(r)-\mathbf{P}\cdot(\bm{\omega}\times\mathbf{r})\nonumber\\
&=\frac{1}{2m}\frac{p^2}{1-\beta p^2} - \frac{1}{8m^3 c^2}\frac{p^4}{(1-\beta p^2)^2}
\nonumber\\
& \hspace{0.4cm} +m\phi(r) - \frac{\mathbf{p}\cdot (\bm{\omega} \times \mathbf{r})}{\sqrt{1-\beta p^2}},
\end{align}
where we have omitted $\mathcal{O}\left(1/c^{4}\right)$ terms. To first order in $\beta$, we obtain
\begin{align}
H^{GUP}_{R, S} &= \frac{p^2}{2m} + \frac{\beta}{2m}p^4 - \frac{p^{4}}{8m^{3}c^{2}}\left(1+2\beta p^2\right)+m\phi(r)
\nonumber\\
&\hspace{0.4cm}-\mathbf{p}\cdot(\bm{\omega \times \mathbf{r}})\left(1+\frac{1}{2}\beta p^2\right),
\end{align}
where we have also omitted $\mathcal{O}\left(\beta^{2}\right)$ terms. The phase can be computed as in the non-relativistic case. The phase shift arising from the terms in $\omega$ reads
\begin{align}\label{eq:GUPR}
\Delta \alpha^{GUP}_{R, S} &=\Gamma(P_0)\Delta\alpha_{S}\left(1+\beta P_0^{2}\right)^{-3/2}\nonumber\\
&\simeq\Gamma(P_0) \Delta \alpha_{S}
\left(1-\frac{3}{2}\beta P_0^{2}\right),
\end{align}
where $\Gamma(P_0)\equiv\sqrt{1+P_0^{2}/(m^{2}c^{2})}$ reduces to $\gamma_0$ for $\beta=0$. In the non-relativistic limit, $c\to\infty$ and $\Gamma(P_0)\to1$, so Eq.~(\ref{eq:GUPR}) reduces to Eq.~(\ref{eq:GUPsag}), as expected.
Comparing with the undeformed phase at the same momentum $P_0$, and using $P_0=\gamma_0mv_0+\mathcal{O}(\beta)$, the relative deviation is
\begin{align}
\left\lvert\frac{\Delta \alpha_{R, S}^{GUP}-\Gamma(P_0)\Delta \alpha_{S}}{\Gamma(P_0)\Delta \alpha_{S}} \right\rvert&=\frac{3}{2}\beta P_0^{2}\nonumber\\
&\simeq\frac{3}{2}\beta m^{2}v_0^{2}\left(1+\frac{v_0^{2}}{c^{2}}\right).
\label{dR}
\end{align}
To this order in rotation, a comparison at fixed wavelength holds $p_0$ fixed instead. The difference affects the relativistic part of the GUP deviation; the rounded estimates below agree, since $v_0^{2}/c^{2}\lesssim1.2\times10^{-3}$ for the speeds considered.

Some comments are in order here. The GUP and the related minimal length are usually formulated in non-relativistic quantum mechanics. Extending them to relativistic theories is important in order to obtain a Lorentz-invariant minimal length. A covariant generalization of the standard GUP has been formulated in Ref.~\cite{Bosso}. Starting from that GUP, one could follow the approach used here to derive the modified Sagnac phase.

\subsection{General-relativistic corrections}
We now assume that the external gravitational field of the Earth is described by the Kerr metric. In the weak-field approximation, retaining the stated post-Newtonian terms and only first order in the Earth's angular momentum, the line element reads~\cite{GR}
\begin{align}
& ds^2 \simeq \left(c^2 + 2\phi + \frac{2\phi^2}{c^2}\right)dt^2
\nonumber\\
&\hspace{0.9cm} - \frac{\phi}{c^2}\frac{4b}{r^2}(x'dy'- y'dx')dt
\nonumber\\
&\hspace{0.9cm} -\left(1-\frac{2\phi}{c^2}\right)(dx'^2+dy'^2+dz'^2),
\end{align}
where $\phi$ is the Newtonian gravitational potential of the Earth, $\phi = -GM/r$, with $r = \sqrt{x'^2 + y'^2 + z'^2}$, and $\mathbf{b}$ is the angular momentum per unit mass of the Earth, $\mathbf{b} = \mathbf{J}/M = \frac{2}{5}R^{2}\bm{\omega}$, where we have modelled the Earth as a homogeneous sphere of radius $R$. In order to express physical quantities in an Earth-centred frame co-rotating with the Earth, with coordinates $(x,y,z)$, we perform the transformation
\begin{equation}
\begin{array}{l}
t' = t,\qquad z' = z,\\
x' = x\cos{(\omega t)} - y\sin{(\omega t)},\\
y' = x\sin{(\omega t)} + y\cos{(\omega t)}.\\
\end{array}
\end{equation}
Following Ref.~\cite{GR2}, we use the 3+1 formalism to describe the spacetime, where the four-dimensional metric tensor $g_{\mu \nu}$ is split as follows:
\begin{equation}
\begin{array}{l}
g_{00} = N^{2} - \gamma_{ij}N^{i}N^{j},\\
g_{0i} = -\gamma_{ij}N^{j},\\
g_{ij} = -\gamma_{ij},\\
\end{array}
\end{equation}
where $N$ is the lapse function, $N^i$ is the shift vector, and $\gamma_{ij}$ denotes the spatial metric on the hypersurfaces $t=$ const. The slowly rotating, weak gravitational field of the Earth in these co-rotating Earth-centred coordinates is described by
\begin{equation}\label{eq:32}
\begin{array}{l}
\vspace{0.1cm}N = c\left(1 + \frac{\phi}{c^2}+\frac{\phi^2}{2c^4}\right),\\ \vspace{0.1cm}
N^x=-\left[1+\frac{4\phi R^2}{5c^2 r^2}\right]\omega y,\\
N^y=\left[1+\frac{4\phi R^2}{5c^2 r^2}\right]\omega x,\\
N^z = 0,\\
\gamma_{ij} = \left(1-\frac{2\phi}{c^2}\right)\delta_{ij}.\\
\end{array}
\end{equation}
In order to derive the Hamiltonian of a free neutron moving in the curved space-time produced by the Earth, we consider the relativistic Lagrangian of a particle of mass $m$~\cite{GR3}:
\begin{align}
L_{GR} &= -mc\sqrt{g_{\mu \nu}\dot{x}^\mu \dot{x}^\nu}
\nonumber\\
&= -mc\sqrt{N^2 - \gamma_{ij}(N^i + \dot{x}^i)(N^j + \dot{x}^j)}.
\end{align}
Using the canonical momentum $p_{i}=\partial L_{GR}/\partial\dot{x}^{i}$, we write down the Hamiltonian as
\begin{eqnarray}
H_{GR} &\equiv& p_i\dot{x}^i - L_{GR} - mc^2 \nonumber\\
&=& N\sqrt{m^2 c^2 + \gamma^{ij}p_{i}p_{j}} - N^i p_i -mc^2,
\end{eqnarray}
where the rest-mass energy has been subtracted from the conventional definition of the Hamiltonian for later convenience. In the frame centred on and rotating with the Earth, we replace the canonical momentum $p_i$ with the deformed momentum $P_i$. The Hamiltonian becomes
\begin{equation}\label{eq:GUPGR}
H^{GUP}_{GR}=N\sqrt{m^2c^2+\gamma^{ij}P_iP_j}-N^iP_i-mc^2,
\end{equation}
where
\begin{equation}
P_i=\frac{p_i}{\sqrt{1-\beta \gamma^{kl}p_k p_l}}.
\end{equation}
By defining $p_{\gamma}^2 \equiv \gamma^{ij}p_i p_j$, the Hamiltonian~(\ref{eq:GUPGR}) becomes, to first order in $\beta$,
\begin{align}
H^{GUP}_{GR}&=  N\sqrt{m^2c^2+p_{\gamma}^2(1+\beta
p_{\gamma}^2)}\nonumber\\
&\hspace{0.4cm}-N^ip_i\left(1+\frac{1}{2}\beta p_{\gamma}^2\right)-mc^2,
\end{align}
where we have omitted $\mathcal{O}\left(\beta^{2}\right)$ terms. Now, assuming $p_{\gamma}^{2} \ll m^2 c^2$, expanding up to $\mathcal{O}(1/c^{2})$, and neglecting, here and in the following, terms of order $\beta/c^{2}$, i.e., products of the GUP correction with post-Newtonian corrections (which are suppressed by a further factor $|\phi|/c^{2}\sim10^{-9}$ or $p^{2}/(m^{2}c^{2})$), we get
\begin{align}\label{eq:40}
H^{GUP}_{GR}&= \left(\frac{N}{c}-1\right)mc^2+\frac{N}{c}\frac{p_{\gamma}^2}{2m}(1+\beta
p_{\gamma}^2)\nonumber\\
&\hspace{0.4cm}-\frac{N}{c}\frac{p_{\gamma}^{4}}{8m^{3}c^{2}}-N^ip_i\left(1+\frac{1}{2}\beta p_{\gamma}^2\right)\nonumber\\
&\hspace{0.4cm}+\mathcal{O}\left(\beta/c^{2},1/c^{4}\right).
\end{align}
The above Hamiltonian can be rewritten as follows. First, $N^{i}p_{i}$ can be written as
\begin{equation}
N^ip_i=\left[1+ \frac{4\phi R^2}{5c^2 r^2}\right]\bm{\omega} \cdot \mathbf{L},
\end{equation}
where $\mathbf{L} = \mathbf{r} \times \mathbf{p}$, with $\mathbf{r}=(x,y,z)$ and $\mathbf{p}=(p_{x},p_{y},p_{z})$, and $\bm{\omega}$ is directed along the $z$ axis. The first term on the right-hand side of Eq.~(\ref{eq:40}) can be written as
\begin{equation}
\left(\frac{N}{c}-1\right)mc^2=m\phi+\frac{m}{2c^2}\phi^2.
\end{equation}
Finally, since $\gamma^{ij}=(1+2\phi/c^{2})\delta^{ij}+\mathcal{O}(1/c^{4})$, we have $p_{\gamma}^{2}=(1+2\phi/c^{2})p^{2}$, with $p^{2}\equiv\delta^{ij}p_{i}p_{j}$, and therefore
\begin{align}
\frac{N}{c}\frac{p_{\gamma}^2}{2m}(1+\beta
p_{\gamma}^2)&=\frac{p^2}{2m}+\beta\frac{p^4}{2m}+\frac{3\phi}{c^2}\frac{p^2}{2m}\nonumber\\
&\hspace{0.4cm} + \mathcal{O}\left(\beta/c^{2},1/c^4\right),
\end{align}
where the factor 3 combines the contributions of the lapse function and of the spatial metric. By defining
\begin{equation}
\gamma_G \equiv \frac{4\phi R^2}{5c^2 r^2},
\end{equation}
the Hamiltonian becomes
\begin{align}
H^{GUP}_{GR, S}&=\frac{p^2}{2m}+\beta\frac{p^4}{2m}-\frac{p^{4}}{8m^{3}c^{2}}+m\phi\nonumber\\
&\quad+\frac{3\phi}{c^2}\frac{p^2}{2m}+\frac{m\phi^2}{2c^2}\nonumber\\
&\hspace{0.4cm} -\left(1+\gamma_{G}+\frac{\beta}{2}p^2\right)\bm{\omega}\cdot\mathbf{L}.
\end{align}
\begin{table*}[t]
	\begin{center}
		\caption{Conditional sensitivity projections for horizontal neutron interferometers, using a $10\%$ relative tolerance and the Earth-centred prescription. Each entry assumes $|\bm{\omega}\times\mathbf{r}|\ll v_0$ and the other expansion conditions stated in the text.}
		\label{tab:bounds}
		\vspace{0.5em}
		\renewcommand{\arraystretch}{1.25}
		\begin{tabular}{ccccc}
			\hline\hline
			Sources & $v_0\left(\mathrm{m}/\mathrm{s}\right)$& Eq. & $\beta \left(\mathrm {GeV}^{-2}\right)$& $\beta_0$\\
			\hline
			Cold neutrons & $800$ &\ref{dNR} & $\lesssim 1.1\times 10^{10}$ & $\lesssim 1.6\times 10^{48}$ \\
			Thermal neutrons & $2200$ &\ref{dNR} & $\lesssim 1.4 \times 10^{9}$ & $\lesssim 2.1 \times 10^{47}$ \\
			Fission neutrons & $10^{7}$ & \ref{dR} & $\lesssim 6.8\times 10^{1}$ & $\lesssim 1.0\times 10^{40}$ \\
			\hline\hline
		\end{tabular}
	\end{center}
\end{table*}

At fixed energy, the rotational change in the momentum component along the path is $\delta p=m\,\hat{\mathbf{n}}\cdot(\bm{\omega}\times\mathbf{r})f(\mathbf{r})$, with
\begin{equation}\label{eq:46}
f(\mathbf{r})=\gamma_0-\frac32\beta p_0^{2}-\frac{3\phi(r)}{c^{2}}+\gamma_{G}(r),
\end{equation}
where $\gamma_0\simeq1+p_0^{2}/(2m^{2}c^{2})$ to the retained order. For this small horizontal interferometer, the radial-gradient contribution to the loop integral vanishes at leading order in $\ell/R$. Thus,
\begin{equation}\label{eq:50}
\Delta\alpha_{GR,S}^{GUP}=\Delta\alpha_{S}
\left(\gamma_0-\frac32\beta p_0^{2}-\frac{3\phi}{c^{2}}+\gamma_{G}\right),
\end{equation}
where \(p_0\), \(\gamma_0\), \(\phi\), and \(\gamma_G\) are evaluated at the interferometer centre, consistently with the small horizontal geometry assumed above. In the non-relativistic limit, this reduces to Eq.~(\ref{eq:GUPsag}). For other orientations, the loop integral requires the spatial gradients of the momentum and metric.

Denoting the general-relativistic phase for $\beta=0$ by $\Delta\alpha_{GR,S}$, the leading relative deviation for the same reference beam conditions is
\begin{align}\label{dGR}
&\left|\frac{\Delta\alpha_{GR,S}^{GUP}-\Delta\alpha_{GR,S}}{\Delta\alpha_{GR,S}}\right|\nonumber\\
&\qquad=\frac32\beta m^{2}v_0^{2}
\left[1+\mathcal{O}\left(\frac{v_0^{2}}{c^{2}},\frac{|\phi|}{c^{2}}\right)\right].
\end{align}
The reference phase must dominate the neglected contributions. Mixed terms of order $\beta/c^{2}$ are required to determine the subleading coefficients; at leading order, the deviation agrees with Eq.~(\ref{dNR}). Coordinate quantities must be converted to locally measured ones for a comparison at post-Newtonian accuracy.

\section{Bounds on $\beta$ and concluding remarks}
The COW measurements~\cite{accuracy} have statistical uncertainties of order $10^{-3}$ and residual discrepancies of order $10^{-2}$~\cite{kindsexp}. We use the latter as an illustrative tolerance $\delta$ to estimate sensitivity:
\begin{equation}\label{condition}
\left|\frac{\Delta\Phi^{GUP}-\Delta\Phi}{\Delta\Phi}\right|\lesssim\delta.
\end{equation}
For the non-rotating COW component, Eq.~(\ref{dCOW}) gives $2\beta m^{2}v^{2}\lesssim10^{-2}$ at fixed incident wavelength. With $m=1.675\times10^{-27}\,\mathrm{kg}$ and $v\simeq2.2\times10^{3}\,\mathrm{m/s}$, this corresponds to $\beta\lesssim1.1\times10^{8}\,\mathrm{GeV}^{-2}$ and $\beta_{0}\lesssim1.6\times10^{46}$. Here $\beta_{0}=\beta M_{p}^{2}c^{2}$, with $M_{p}c^{2}=1.22\times10^{19}\,\mathrm{GeV}$; numerical values of $\beta$ in $\mathrm{GeV}^{-2}$ use $\hbar=c=1$. Applying this estimate to a rotating apparatus requires accounting for its rotational phase.

For the neutron Sagnac effect, agreement with theory at the level of a few percent~\cite{NSagnac,sag4,kindsexp} motivates the illustrative tolerance $\delta=0.1$. Table~\ref{tab:bounds} uses Eqs.~(\ref{dNR}) and~(\ref{dR}) for horizontal interferometers, with representative neutron speeds~\cite{kindsexp} assigned to $v_0$. These projections assume the geometry and expansion conditions stated above. In the Earth-centred prescription, $|\bm{\omega}\times\mathbf{r}|$ can be as large as $465\,\mathrm{m/s}$; hence the rotation expansion must be checked for the actual location and speed. The cold and thermal projections therefore remain conditional on the experimental latitude satisfying $|\bm{\omega}\times\mathbf{r}|\ll v_0$.

For these speeds, the largest kinetic relativistic correction is $\gamma_0-1\simeq6\times10^{-4}$, and the gravitational post-Newtonian corrections are of order $10^{-9}$, both below the adopted tolerance. These sensitivity estimates are consistent with, although weaker than, the strongest existing bounds (e.g., $\beta_{0}<10^{36}$ from the Lamb shift~\cite{boundsGUP}); comparisons should account for the different models and normalizations of $\beta$.

Matter-wave interferometry provides a way to test minimal-length models. We have shown that, within the phenomenological Earth-centred prescription considered here, (i) neglecting rotation and to first order in the gravitational potential difference, the COW phase has its standard velocity form but acquires a relative correction $-2\beta p^{2}$ at fixed wavelength; (ii) for a small horizontal interferometer, the leading Sagnac phase acquires a relative correction $-\frac32\beta p_0^{2}$; and (iii) relativistic effects enter the GUP deviation at subleading order in this geometry. We have used these results to give conditional sensitivity estimates for the dimensionless GUP parameter $\beta_{0}$.


\end{document}